\documentclass[namedreferences]{solarphysics}
\usepackage[hyperref,optionalrh,solaromanenum]{spr-sola-addons} 
\usepackage{graphicx}                    
\usepackage{color}                       
\usepackage{breakurl}                         
\usepackage{multirow}

\chardef\us=`\_

\begin{document}

\begin{article}

  \begin{opening}

    \title{Temporal and Periodic Variations of Sunspot Counts in Flaring and Non-flaring Active Regions}

    \author[addressref={aff1},corref,email={alikilcik@akdeniz.edu.tr}]{\inits{A.}\fnm{A.}~\lnm{Kilcik}}
    \author[addressref={aff2,aff3},{email=vayur@bbso.njit.edu}]{\inits{V.}\fnm{V.}~\lnm{Yurchyshyn}}
    \author[addressref=aff1,email={burcindonmez@akdeniz.edu.tr}]{\inits{B.}\fnm{B.}~\lnm{Donmez}}
    \author[addressref=aff4,email={obridko@mail.ru}]{\inits{V.N.}\fnm{V.N.}~\lnm{Obridko}}
    \author[addressref=aff5,email={ozguc@boun.edu.tr}]{\inits{A.}\fnm{A.}~\lnm{Ozguc}}
    \author[addressref=aff6,email={jp.rozelot@orange.fr}]{\inits{J.P.}\fnm{J.P.}~\lnm{Rozelot}}

    \address[id=aff1]{Akdeniz University Faculty of Science, Department of Space Science and Technologies, 07058, Antalya, Turkey}
    \address[id=aff2]{Big Bear Solar Observatory, Big Bear City, CA 92314, USA}
    \address[id=aff3]{Korea Astronomy and Space Science Institute, 776 Daedeok-daero, Yuseong-gu, Daejeon, 305-348, South Korea}
    \address[id=aff4]{Pushkov Institute of Terrestrial Magnetism, Ionosphere and Radio Wave Propagation of the Russian Academy of Sciences
    	(IZMIRAN), Troitsk, Moscow, 142190 Russia}
    \address[id=aff5]{Kandilli Observatory and Earthquake Research Institute, Bogazici University, 34684 Istanbul, Turkey}
    \address[id=aff6]{Universit\'{e} de la C\^{o}te  d'Azur (OCA-CNRS) and 77, Ch. des basses Moulieres, 06130 Grasse (F)}

    \runningauthor{Kilcik et al.}
    \runningtitle{Comparison of Flaring and non-flaring ARs}

    \begin{abstract}
      We analyzed temporal and periodic behavior of sunspot counts (SSCs) in flaring (C, M, or X class flares),  and
      non-flaring active regions (ARs) for the almost two solar cycles (1996 through 2016). Our main findings 
      are as follows: i) The temporal variation of monthly means of daily total SSCs 
      in flaring and non-flaring ARs are different and these differences are also varying
      from cycle to cycle; temporal profile of non-flaring ARs are wider than the flaring ones during the solar cycle 23, while they are almost the same during the current cycle 24. 
      The second peak (second maximum) of flaring ARs are strongly dominate during current cycle 24, while this difference is not such a remarkable during cycle 23. The amplitude of SSCs in
      the non-flaring ARs are comparable during the first and second peaks (maxima) of the current solar cycle, while the first peak is almost not existent in case of
      the flaring ARs.
      ii) Periodic variations observed in SSCs of flaring and non-flaring ARs are 
      quite different in both MTM spectrum and wavelet scalograms and these variations are also different from one cycle to another; the largest detected
      period in the flaring ARs is 113 days, while there are much higher periodicities (327, 312, and 256 days) in non-flaring ARs. There are no meaningful periodicities
      in MTM spectrum of flaring ARs exceeding 45 days during solar cycle 24, while a 113 days periodicity detected from flaring ARs of solar cycle 23. For the non-flaring ARs the largest period is 72 days during solar cycle 24, while the largest period is 327 days during current cycle.   
    \end{abstract}

    %
    \keywords{Sun: Active Regions, Sunspots, Flares, Periodicity}

  \end{opening}

  %
   \section{Introduction}\label{s:intro} 
   Our sun is a variable star and it shows different kind of variations such as sunspots, solar flares, prominences, coronal mass ejections etc., as depending on the observed wavelength and depth. These variations can be traced easily by observing such solar activity indicators, as sunspot numbers (SSNs), sunspot areas (SSAs), 10.7 cm solar radio flux (F10.7), solar flare index (FI), total solar irradiance (TSI), etc. All of these indicators describe solar activity quite well. Due to the longest temporal coverage (about 400 years), sunspots are one of the most important and commonly used indicators. Studies on variations in solar activity are important for understanding the mechanism behind the solar activity and solar cycle, and also for predicting the level of activity \citep{Hathaway09, Petrovay10}. The properties of solar cycle are generally described by the Zurich or international sunspot number, RZ = k(10g+f), where k is a correction factor for the observer, g is the number of identified sunspot groups, and f is the number of individual sunspots.  According to this equation daily sunspot number strongly related to the number of observed group for a day. 
   
   In the photosphere the presence of a strong magnetic field is manifested by the appearance of dark sunspots or pores and bright faculae representing concentrated and dispersed magnetic fields, respectively. In general, sunspots are observed as groups on the active regions (ARs) of solar surface and they are classified according to their morphology and evolution. The ARs may be classified in terms of the morphology of the sunspot groups. The most common classification of ARs was introduced by \citet{McIntosh90}. The McIntosh Sunspot Classification Scheme (MSCS) assigns three descriptive codes characterizing the size (A, B, C, D, E, F, H), penumbra (X, R, S, A, H, and K) and compactness (X, O, I, and C) of ARs. The simplest ARs have bipolar magnetic field configurations, but ARs may be built-up by several bipoles emerging in close succession.  In this classification A and B classes describe very tiny small SGs
   and they produce very rare flare, while D, E and F describe strongly evaluated complex ones and most of the flares produced by these evaluated groups. Note that in the calculation of daily sunspot number all SGs have the same weight. 
   
   All solar activity indicators show a periodic behavior with period ranging from days to hundreds of years. The best-known solar periodic variations are the eleven year sunspot cycle and the 27-day solar rotation periodicity that is induced by large long-lived ARs with lifetimes longer than one solar rotation. Investigations of possible periodicities other than these two periods have been of interest for a long time. Let us recall these studies;  in 1984, Rieger \citep{Rieger84} was the first to reveal a 158-day periodicity in the Sun, while studying $\gamma$-ray flare data from Solar Maximum Mission (SMM) in solar cycle 21 (C21). Approximately, the same periodicity was also discovered in X--ray flares data taken from the Geosynchronous Operational Environmental Satellites (GOES) for the same solar cycle \citep{Rieger84}. This was only the beginning in a sequence of relevant investigations. In the context of these attempts, it is notable that apart from the Rieger periodicity itself, numerous other periodicities were discovered, such as 128, 102, 78 and 51-day periodicities \citep{Bai91, Bai92}. Consequently, Rieger periodicity was connected to highly energetic flares which are presumably triggered by the emergence of photospheric magnetic flux with the same period  \citep{Ballester99, Ballester02}. The most important short-time periodicities in different kinds of solar activity, as reported before, are as follows: i) a 153-day period in H$\alpha$ flare importance \citep{Ichimoto85}, hard X--ray peak rate \citep{Dennis85, Bai87, Verma91}, in flare index \citep{Ozguc89, Ozguc94, Ozguc03}, 10.7--cm radio peak flux \citep{Kile91}, and production of energetic electrons in the interplanetary space \citep{Droege90}; ii) 323-- and 540-day periodicities in daily sunspot number and sunspot area \citep{Oliver92} in flare index \citep{Kilcik10}; iii) a quasi--period of about 160 days in the photospheric magnetic flux \citep{Ballester02}; iv) 157$\pm$11 day periodicity in X--ray flares greater than M5 \citep{Lou03}.

   On the other hand, large-scale solar activity is produced by the energy stored active region magnetic fields. Consequently flares often occur at sites in active regions overlying neutral lines where fields are strongly sheared. Observations show that the photospheric topology of the magnetic field is one of the key factors in determining the evolution of ARs. There is a general trend for large ARs to produce large flares and for more complex ARs to generate more numerous and larger flares than other ARs of similar size \citep{Sammis00}. Here, we separated SGs in two categories based on their ARs’ flare productivity: if a sunspot group produced any flare (C, M, and X class flares) during its evolution we count it as a flaring AR, otherwise we count it as a non-flaring AR. Then, we analyzed temporal and periodic behavior of the sunspot counts (SSCs) in these ARs and in an attempt to more comprehensively describe general characteristics of flaring and non-flaring ARs for the last two solar cycles (1996 through 2016). We hope, this study will bring better understanding  of the solar cycle, the background physical phenomena, and consequently it may increase the knowledge about Sun-Earth interactions.
   
   \section{Data and Analysis Method}\label{s:data}
   In this study, we investigate temporal and periodic variations of sunspot counts 
   depending on the flare production of ARs. The group classification 
   and the flare data are only available since August 1996. Therefore, the analyzed time 
   interval covers only solar cycle 23 (1996 through 2008) and the ascending and 
   maximum phases of current cycle 24 (2009 through 2016). The raw data are taken 
   from the Space Weather Prediction Center
   (SWPC)\footnote{ftp://ftp.swpc.noaa.gov/pub/warehouse/}. These data sets include all X-Ray 
   flares and AR informations. The daily total sunspot counts for flaring and non-flaring ARs 
   were calculated. To compare the temporal variations of sunspot counts (SSCs) in 
   flaring and nonflaring ARs for the investigated time period the monthly mean 
   values were calculated. Then, to remove the short term fluctuations and reveal 
   the long term trend a 12 steps running averaging was applied. 

   To reveal periodic variations in the SSC data two period analysis methods, Multitaper Method 
   (MTM) and Morlet Wavelet Analysis, were applied with red noise approximations
   and 95\% confidence level. Both method have previously been successfully used analysis of solar data \citep[][and reference therein]{Prestes06,  
   Kilcik10, Mufti11, Deng13, Choudhary14, Kilcik14b, Kilcik16}.

   To obtain the periodicities with high confidence level we used the MTM analysis method 
   which enables us to detect low-amplitude harmonic oscillations in 
   relatively short time-series with a high degree of statistical significance rejecting 
   larger amplitude harmonics if the F-test (variance test) fails. This 
   feature is an important advantage of the MTM over other classical methods
   (\textit{for details see,} \citep{Jenkins69, Ghil02}). In this study we used three sinusoidal tapers, 
   and the frequency range was chosen from 0.0018 to 0.04 (i.e. 25 -- 546 days). 

   To localize the periodicities obtained from the MTM analysis we applied the Morlet 
   Wavelet analysis method to the daily SSC data separately for flaring and nonflaring ARs. 
   The Morlet wavelet is a complex sine wave, localized within a Gaussian window 
   \citep{Morlet82}. The method is a powerful tool for analyzing localized power 
   variations within a time series \citep{Torrence98}. The standard Interactive 
   Data Language (IDL) packages for Morlet wavelet analysis were used, and the scalograms
   were obtained to study both the presence and evolution of the periodicities.

   \section{Results}\label{s:results}
   %
   \begin{figure} 
     \centerline{\includegraphics[width=0.7\textwidth]{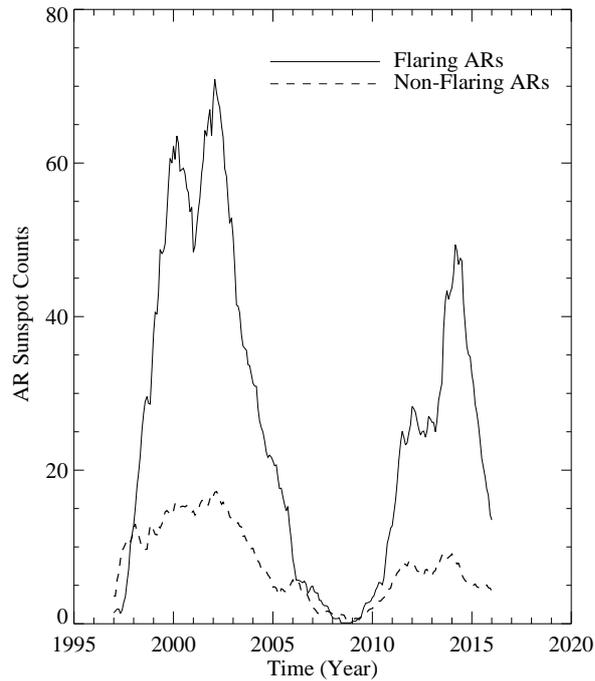}}
     \caption{Temporal variations of SSCs in flaring (solid line) and non-flaring 
     (dashed line) ARs for 1996--2016 time period.}\label{f:fig1}
   \end{figure}

   Figure \ref{f:fig1} shows the temporal variations of SSCs in flaring and non-flaring 
   ARs for the investigated time interval (from 1996 to 2016).  In general, the monthly 
   means of the daily total SSCs in the flaring ARs are four times larger than the non-flaring ones.
   We emphasize that the SSCs in both data sets follows sunspot cycle with some
   differences: First, SSCs in the non-flaring ARs show a wider temporal profile than that of the
   flaring ARs. Second, SSCs in the non-flaring ARs have multiple peaks during the 
   maximum phase of solar cycle 23, while the flaring ARs have two prominent peaks at the 
   same phase. Third, SSCs in the non-flaring ARs have comparable magnitude of the first and 
   second peaks in the current solar cycle, while the first peak is almost 
   not existent in case of the flaring ARs. Finally, there is a well defined peak in SSCs of 
   the non-flaring ARs during 2006--2007, while that peak is not so prominent in case of the flaring ARs. 

   %
   \begin{figure} 
     \centerline{\includegraphics[width=0.75\textwidth]{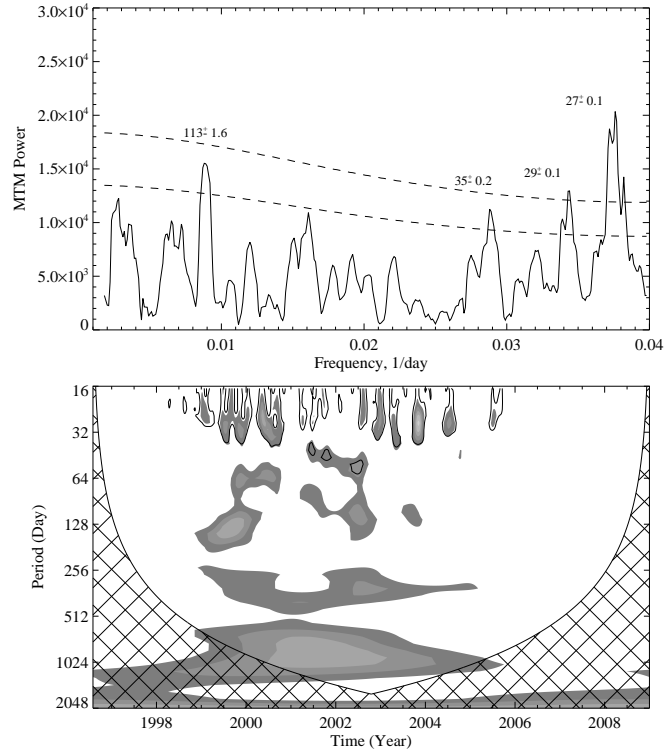}}
     \caption{The multi-taper method (MTM, upper panel) and the Morlet wavelet
     (lower panel) analysis results for SSCs in flaring ARs during the solar Cycle 23
     (1996--2008). The main peaks are labeled with the period in days in the MTM power spectrum. The horizontal dashed lines indicate the 95\% and 99\% 
     confidence levels.The black contours in the wavelet scalogram indicate the 95\% 
     confidence level and the hatched area below the thin black line is the cone 
     of influence (COI).}\label{f:fig2}
   \end{figure}

   %
   \begin{figure} 
     \centerline{\includegraphics[width=0.75\textwidth]{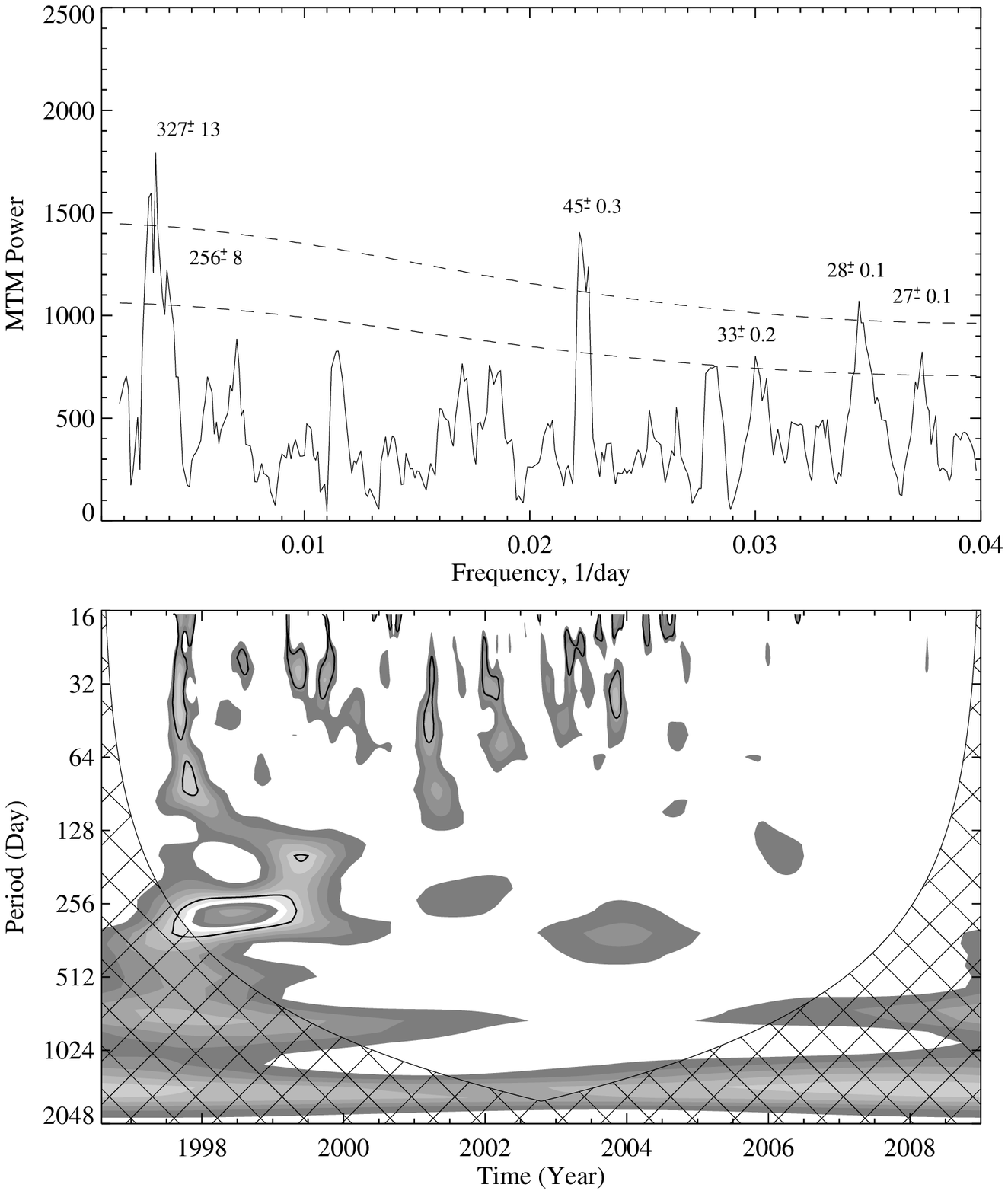}}
     \caption{Same plot as in Figure \ref{f:fig2} for non-flaring ARs.}\label{f:fig3}
   \end{figure}

   Figure \ref{f:fig2} and \ref{f:fig3} show observed periods of SSCs in flaring and non-flaring ARs   
   and their localizations during the solar cycle 23. 
   Detected periods are quite different in both MTM spectrum and wavelet 
   scalograms: i) the largest detected period in the flaring ARs is 113 days, 
   while there are much higher periodicities (327, 312, and 256 days) in 
   non-flaring ARs, ii) there is a very prominent 45 day peak in non-flaring 
   ARs, which does not appear in flaring ARs. The only similarity 
   between the periodicities of the two data sets is the existence of solar 
   rotation periods (27-–35 days). We note that all periods detected with the MTM 
   appear in the wavelet scalograms, but significance level of some periodicities is below the 95\% confidence level. 

   %
   \begin{figure} 
     \centerline{\includegraphics[width=0.75\textwidth]{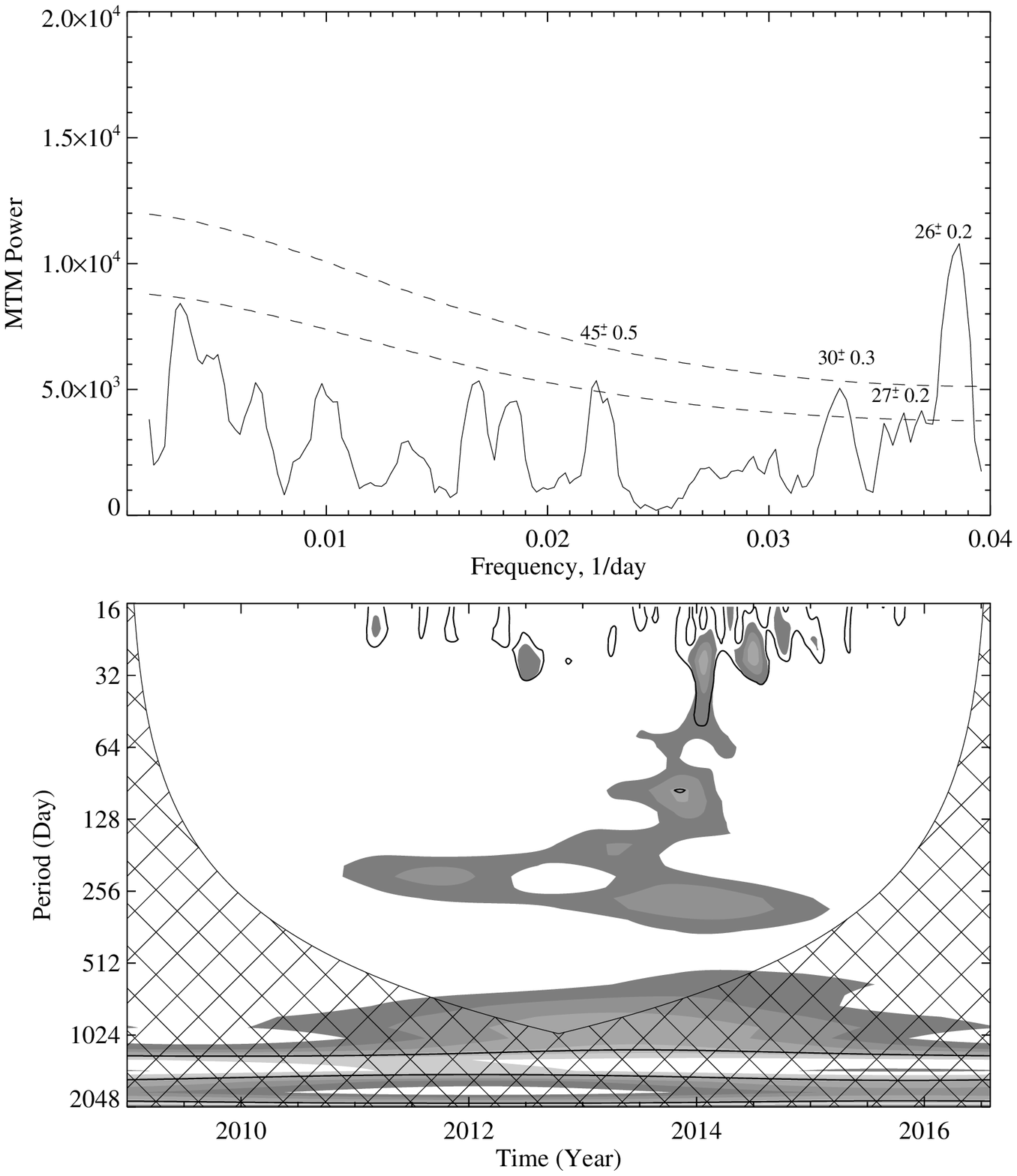}}
     \caption{The multi-taper method (MTM, upper panel) and the Morlet wavelet
     	(lower panel) analysis results for SSCs in flaring ARs during the solar Cycle 24
     	(2009--2016). The main peaks are labeled with the period in days in the MTM power spectrum. The horizontal dashed lines indicate the 95\% and 99\% 
     	confidence levels.The black contours in the wavelet scalogram indicate the 95\% 
     	confidence level and the hatched area below the thin black line is the cone 
     	of influence (COI).} \label{f:fig4}
   \end{figure}

   %
   \begin{figure} 
     \centerline{\includegraphics[width=0.75\textwidth]{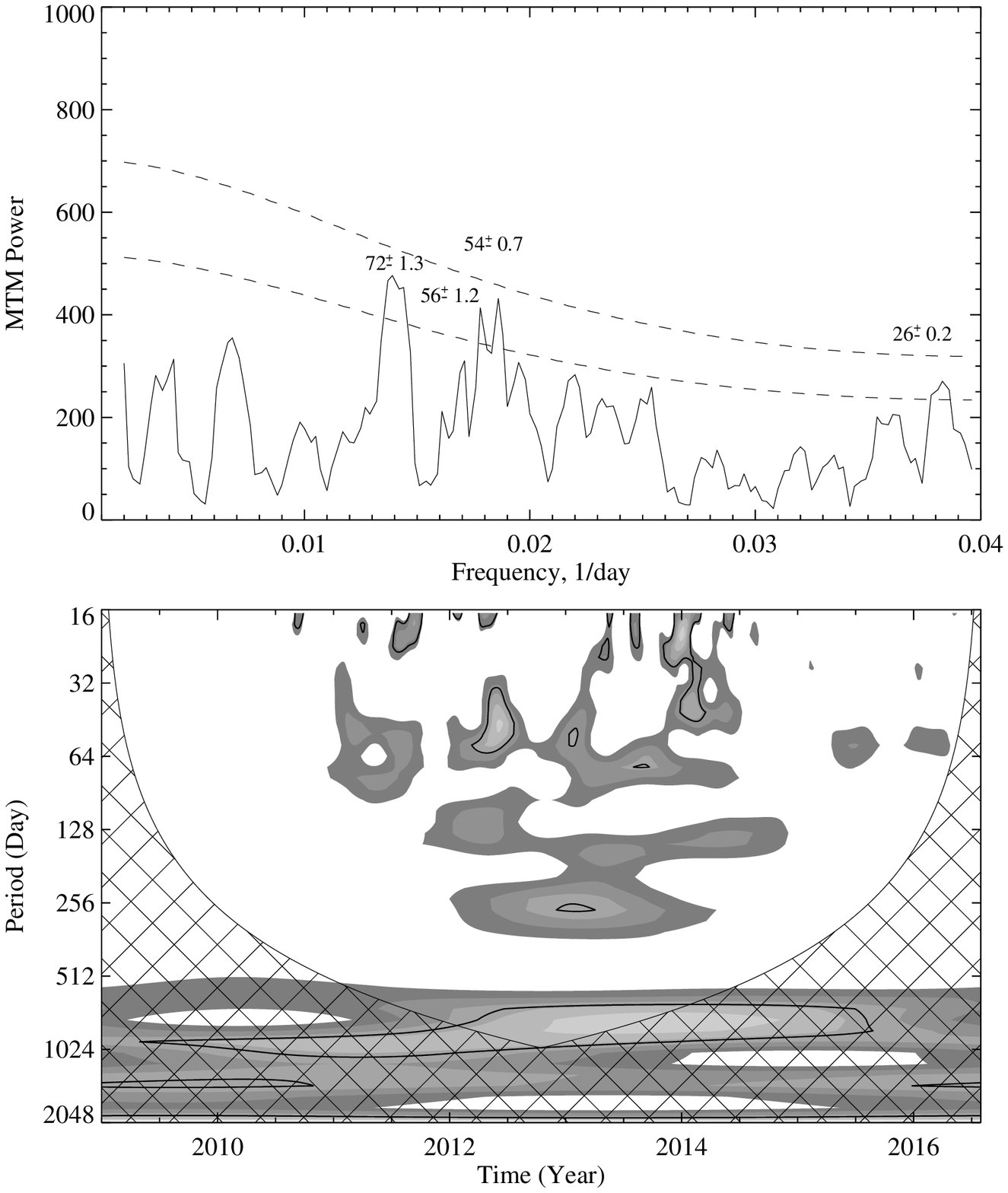}}
     \caption{Same plot as in Figure \ref{f:fig4} for non-flaring ARs.}\label{f:fig5}
   \end{figure}

   Figure \ref{f:fig4} and \ref{f:fig5} show the SSCs periodic variations 
   for the ascending and the maximum phases of solar cycle 24 which also show remarkable differences 
   between the flaring and non-flaring ARs: There are no 
   meaningful periodicities in MTM spectrum of flaring ARs exceeding 45 days, while we detect 72 
   day periodicity in non-flaring ARs. Note that some larger periodicities (about 250 
   and 600 days) appear in the wavelet scalograms of non-flaring ARs. Again solar rotation 
   periodicities exist in both cases.   

   We also noted that in general, the appearance of all well defined periodicities (except the solar rotation period) 
   are shifted toward the late phase of both cycle wavelet scalograms of the flaring 
   ARs, while the non-flaring ARs do not show such a shift.  

   In Table \ref{t:tbl1}, we list significant periods that appeared in 
   flaring and non-flaring ARs in solar cycles 23 and 24. The table 
   shows five significant points: i) large periods (\textgreater 113 day) exist only in the SSCs of 
   non-flaring ARs, ii) a 113 day periodicity appear only in flaring ARs of solar cycle 23, 
   iii) 72 and 55 days periods exist only in the non-flaring ARs 
   of current cycle 24, iv) a 45 day periodicity appear only in flaring ARs of 
   both solar cycles (Cycles 23 and 24), v) 26--35 days solar rotation 
   periodicity exist in all cases. 

   %
   \begin{table}
     \caption{Periods obtained from the MTM analysis. The first column corresponds to the 
     obtained periods and the rest of the columns show the presence of these periods 
     in all cases by means of their significance levels.}\label{t:tbl1}
     \begin{tabular}{ccccc}
       \noalign{\hrule height 1.2pt}
       \multirow{2}{*}{Period} & \multicolumn{2}{c}{Flaring ARs} &
       \multicolumn{2}{c}{Non-flaring ARs} \\  \cline{2-5}
                & Cycle 23 & Cycle 24 & Cycle 23 & Cycle 24   \\ \noalign{\hrule height 1.2pt}
       315--327 & --       & --       & $+>$95   & --         \\ \hline
       293      & --       & --       & $+>$95   & --         \\ \hline
       256      & --       & --       & $+>$95   & --         \\ \hline
       113      & $+>$95   & --       & --       & --         \\ \hline
       72       & --       & --       & --       & $+>$95     \\ \hline
       54--56   & --       & --       & --       & $+>$95     \\ \hline
       45       & $+>$99   & $+>$95   & --       & --         \\ \hline
       26--63   & $+>$95   & $+>$95   & $+>$95   & $+>$95     \\ \noalign{\hrule height 1.2pt}
     \end{tabular}
   \end{table}

   \section{Conclusions and Discussions}\label{s:conc}
   Here we analyzed temporal and periodic behavior of SSCs in flaring and 
   non-flaring ARs and in an attempts to better explain different behaviors of these 
   ARs for the last two solar cycles (1996 through 2016). We have two main 
   findings.

   \begin{enumerate}
     \item The temporal variation of monthly means of daily total SSCs 
     in flaring and non-flaring ARs are different and these differences are also varying
     from cycle to cycle; temporal profile of non-flaring ARs are wider than the flaring ones during the solar cycle 23, while they are almost the same during the current cycle 24. 
     The second peak of flaring ARs are strongly dominate during current cycle 24, while this difference is not such a remarkable during solar cycle 23. SSCs in
     the non-flaring ARs have comparable magnitude of the first and second peaks in the current solar cycle, while the first peak is almost not existent in case of
     the flaring ARs
    
     \item Periodic variations observed in SSCs of flaring and non-flaring ARs are 
     quite different in both MTM spectrum and wavelet scalograms and these variations are also different from cycle to cycle; the largest detected
     period in the flaring ARs is 113 days, while there are much higher periodicities (327, 312, and 256 days) in non-flaring ARs. There are no meaningful periodicities
     in MTM spectrum of flaring ARs exceeding 45 days during current cycle 24, while a 113 days periodicity detected from flaring ARs of solar cycle 23. For the non-flaring ARs the largest period is 72 days during current cycle 24, while the largest period is 327 days during solar cycle 23.  
   \end{enumerate}  

   \subsection{Temporal variation of flaring and non-flaring active regions}
   In the recent years, many studies were focused on the understanding of temporal 
   variation of solar activity indicators (e.g. SSCs, SSAs, Solar X-Ray flares) 
   in various categories and found that these variations change during a solar cycle and from one cycle to another. 
   \citet{Kilcik11} first separated sunspot groups in two categories as small
   and large and analyzed the temporal variation of the SG numbers for the last 
   four solar cycles (Cycles 20, 21, 22 and 23). They conclude that these two 
   categories behave differently during a cycle. In the following years, these 
   different behavior of sunspot groups are confirmed by different authors from 
   different solar activity indicator data sets \citep[][and reference therein]{Lefevre11, Gao12, 
   Nagovitsyn12, Javaraiah13, Gomez14, Kilcik14a, Kilcik14b, GaoZ16}. Here, we separated sunspot groups in two categories 
   as flaring and non-flaring ARs according to their flare production. Then, 
   we investigated the temporal variations of sunspot counts in these ARs for 
   the last two solar cycles. We found that flaring and non-flaring ARs behave 
   differently for different cycles; flaring ARs have remarkable peaks at the second peaks of both solar cycle 23 and 24, while these two peaks have almost the same amplitude in case of the non-flaring ARs during the both cycles. SSCs in the non-flaring ARs show a wider temporal profile than that of the flaring ARs. Also SSCs in the non-flaring ARs have multiple peaks during the maximum phase of solar cycle 23, while
   the flaring ARs have two prominent peaks at the same phase.These results show that the flaring ARs are better describe the large sunspot groups compared to non-flaring ones. Thus, we confirm that 
   the most of the flare activity occur in large and complex SGs and further conclude 
   that the temporal variations of flaring and non-flaring ARs are quite different.

    \subsection{Periodic variation of flaring and non-flaring active regions}
   \citet{Lean89} found a 323 day periodicity in various solar 
   activity indicators which were the sunspot-blocking function, Zurich sunspot
   numbers, F10.7, and CaII K plage index. They concluded that this periodicity 
   can have a real solar origin. Later \citet{Kilcik14a} separated sunspot 
   groups in four categories as small (A and B), medium (C), large (D, E, F) and 
   final (H Modified Zurich Classes) and investigated the periodic variation of 
   SSCs in these categories. They found this periodicity as 315 -- 348 days in all 
   categories except simple ones (A and B Zurich classes). Recently, \citet{Kilcik16} 
   analyzed periodicities of the central latitude of SSGs (active latitude) and found
   about 300-370 days periodicities by using of MTM and the Morlet wavelet analysis 
   techniques for the South hemisphere of solar cycle 18, the North hemisphere during solar 
   cycle 20 and both North and South hemispheres of solar cycle 23. Here we 
   analyzed SSCs periodicities in flaring and non-flaring ARs and found that 
   this periodicity exist only in SSCs of non-flaring ARs of cycle 23. 

   293 and 256 days periodicities found in this study were also reported 
   previously by different authors \citep[][and reference therein]{Sello03, Lou03, Kilcik10}. 
   \citet{Chowdhury09} have reported a 261 day 
   period from sunspot areas by means of Scargle periodogram and Morlet wavelet 
   transform for cycles 22 and 23. \citet{Lou03} have found these periods as 
   259.48 $\pm$24.23 days by applying the Fourier power spectral analysis to X-ray
   flares exceeding M5 class during the maximum phase of Cycle 23 (1999 -– 2003), 
   \citet{Scafetta13} were also found a 293 day periodicity from the TSI 
   data by using a multiscale dynamical spectral analysis technique from 2003.15 
   to 2013.16 (descending phase of Cycle 23 and ascending phase of Cycle 24). 
   Here we found that these two periods also exist in non-flaring active regions
   of solar cycle 23. Note that about 256 days period appear in the wavelet 
   scalogram of SSCs in non-flaring ARs of current cycle 24, but its significance
   level is below the 95\% confidence level in the MTM spectrum. Thus we 
   may suggest that these large periodicities may arise from the non-flaring well developed SGs.  

   \citet{Lou03} found 122.19$\pm$4.88 day periodicity by means of 
   the Fourier power spectra analysis of X-ray flares exceeding M5 class during 
   the maximum phase of cycle 23 (1999 –- 2003). \citet{Ozguc02} 
   have investigated the periodicity of solar H-alpha flare index during the 
   ascending branch and the maximum phase of solar cycle 23, and reported that 
   the 116- and 125-day periodicities are in operation during the investigated
   time period. Later, \citet{Kilcik10} analyzed H-alpha solar flare index 
   data by using MTM and Morlet wavelet analysis techniques for 1976-2007 time 
   intervals. They investigated the cyclic variation of whole data and each 
   cycle (Cycle 21, 22, and 23) separately and found a 113 day period from 
   whole and cycle 23 data. Later, \citet{Kilcik14b} found this period in the 
   SSCs of the simple and final groups. We found that this periodicity only 
   appears in the flaring ARs of solar cycle 23. Thus, we confirm above results 
   and further argue that this periodicity is one of the characteristic 
   periodicity of solar flares and also flaring ARs.

   We detected 72 and 54--56 day periods from only non-flaring ARs of current 
   cycle 24. Those periods are also reported in the literature by different
   authors \citep[][and reference therein]{Zieba01, Bai03, Ozguc03, Dimitripoulou08, Kilcik10, Chowdhury13, 
   Kilcik14b, Kilcik16}. 
   \citet{Kilcik10} analyzed the periodicity of H-alpha flare index data for the solar cycle 21, 
   22, and 23. They detected these periodicities from solar cycles 21 and 22,
   but not from solar cycle 23 data. \citet{Scafetta13}analyzed the
   TSI data by using a multiscale dynamical spectral analysis technique from 
   2003.15 to 2013.16 (descending phase of Cycle 23 and ascending phase of Cycle 24) 
   and detected a 73 day period. Recently, \citet{Kilcik14b} analyzed the periodic
   variations of SSCs in different categories and found these two periodicities; a 
   73 day periodicity found in all categories except final (H groups) ones, and a 
   53 day periodicity detected in all categories. Note that about 55 day 
   period exist in all situations but their significance level just below the
   95\% confidence level, also it appears in the wavelet scalograms.

   We detect a 45 day period from SSCs of flaring ARs of both solar cycle 
   (Cycle 23 and 24). This periodicity also reported previously by different 
   authors \citep[][and reference therein]{Bai03, Lou03, Lara08, Dimitripoulou08, 
   Kilcik10, Kilcik14b, Kilcik16}. 
   \citet{Lou03} investigate periodicities of X-Ray solar flares by using Fourier power 
   spectral analyses method and Morlet wavelet analysis from 1999 to 2003 and 
   found about 42 days period. Later, \citet{Lara08} investigate the short 
   term periodicity of number of coronal mass ejection by means of the maximum 
   entropy method (MEM) and wavelet analysis during solar cycle 23 and found 
   about 45 day period. Recently, \citet{Chowdhury15} studied quasi-periodic 
   variations of SSA/SSN, 10.7 cm solar radio flux, average photospheric magnetic 
   flux by means of Lomb-Scargle periodogram and Morlet wavelet analyses during 
   solar cycle 24 (from 2009 January to 2013 August). They detected 43 and 45 
   day periods from SSA and SSNs and 45 day period from 10.7 cm solar radio flux, 
   but not in the average photospheric magnetic flux data. We confirm above 
   results and further found that a 45 day period is only appearing in the 
   flaring ARs data of both solar cycles (Cycles 23 and 24). Thus we may 
   conclude that this period is one of the fundamental periods of only flaring 
   ARs and it may be used as a separator between flaring and non-flaring ARs. 

   26--35 days solar rotation periodicity reported in the literature by many authors 
   from different solar activity indicators \citep[][and reference therein]{Bouwer92, Bai03,
   Kilcik10, Chowdhury11, Kilcik14b, Kilcik16}. This periodicity does not show any AR preference. It can be 
   expected because the life time of many evaluated SGs are longer than
   one solar rotation. Therefore those SGs may appear longer than one solar rotation and affect the
   observed daily SSCs. Slight variations could also occur if the active region had a slow longitudinal movement
   ahead of, or lagging behind the rotation.

   \subsection{Physical interpretation}   
   \citet{Parker55}'s dynamo equations, which can be rigorously derived from the general 
   mean-field dynamo equations are today the most common scheme of magnetic field 
   generation \citep{Fioc96, Krause80}. In this scheme, the dipole magnetic field is transformed by the differential rotation into near-equatorial 
   local field of active regions ($\Omega$ branch dynamo). At the second step the polar 
   field is restored from the local  fields (branch $\alpha$). This assumes one cell of 
   meridional circulation. Field transport from the pole to the equator occurs at
   the lower boundary of the convection zone (tachocline), while the movement from 
   equator to pole just below the surface. 

   This is a greatly simplified scheme and is currently undergoing considerable 
   criticism and is complemented by a variety of details. In particular, it is unclear 
   how magnetic field generated at the base of the convection zone rises up, and in which form it is made available to the observations at the 
   photosphere level. Nevertheless, an important argument in favor of this branch of the
   dynamo is the high correlation between the polar field at the minimum  of the cycle and 
   the next maximum \citep{Obridko08}.

   But why change the polar field? The process of restoration of the polar field to 
   the end is not clear. The most common mechanism is of Babcock-Leighton  
   \citep{Babcock61, Leighton69, Upton14}. However still unclear, 
   what is the ratio of two processes - advection and diffusion
   \citep{Georgieva11, Obridko12a}. 

   Some of the difficulties in the mechanism of Babcock-Leighton are that the total 
   magnetic flux of sunspots varies over the 11-year cycle of 10-12 times, while the 
   large-scale flux  - less than 2 times.  Moreover, the total magnetic flux of sunspots 
   equal not more than 11-14\% of the total solar magnetic flux \citep{Harvey96}. In addition, 
   the meridional flow greatly vary over the cycle, and perhaps are not the cause but the 
   consequence of wave activity. 

  The simplest scheme suppose that the  generated magnetic field emerge from the depth
  to the surface in the form of tubes. However, modes of  active regions appearance 
  do not look like something that should be as a floating monolith tubes. We observe 
  clearly picking up the individual cells of the magnetic field to the general structure 
  \citep{Getling15, Getling16}. Thus one cannot  directly identify the magnetic flux that 
  is generated at great depths, and the number and size of active regions, although 
  they are of course linked.

  At present, already there is a lot of evidence to suggest that there should be, an 
  additional process that converts the generated at greater depths large scale magnetic 
  field in the field of sunspots. This could be distributed dynamo, subsurface dynamo, 
  small-scale dynamo. There are reasons to believe that in the convective zone, 
  there are not one but two or even three cells of  generations 
  \citep{Pipin14, PipinK14, PipinK15, Du15, Yadav15}.
  Apparently, there are other processes that collect emerged weak field on the surface to 
  strong fields of active regions  
  \citep{Getling15, Getling16, Brandenburg16}.

  As a result of  the presence of two related but different processes there are  two important 
  conclusions. Firstly,  because of the large number of external parameters difficult to 
  expect a complete coincidence of two cycles. Initial generation of the magnetic field 
  provides a relatively stable 11- year variation. However, the dependence of the internal 
  activity on the phase of the cycle can vary greatly. This also applies to the 
  spectral composition of solar activity variations. On the one hand, you can expect 
  the existence of a rich spectrum of periodicity, the individual frequencies can appear 
  and disappear at different phases of the cycle. This is what is observed in results of this paper.

  Second, it was to be expected that there are two populations of sunspots. This conclusion is 
  now fully confirmed \citep{Bludova14, Obridko14, Nagovitsyn16}. 
  Populations of large and small groups have different statistical properties. The non-flaring 
  active regions correspond as a rule  to the small size of AR and determined by the primary 
  dynamo mechanism. That is why the hump was observed during the declining phase of solar cycle 23 (in 2005-2006). 
  This maximum was observed by us earlier, and its nature is studied in detail in the paper 
  (\citet{Obridko12b}, see Fig.1).

  By virtue of wealth and temporal instability of the spectrum does not make sense to 
  discuss the nature of each peak in the spectrum. It can be noted that the periods are 
  divided into two groups. The first group  periods around 1 year undoubtedly associated 
  with large-scale fields, they seem connected with quasi-biennial oscillations (QBO). 
  It is known that QBO have a fine structure and essentially represent a set of pulses with 
  a typical repetition rate of 1 to 3 years. This whole set periods could not be reflected 
  in the observed spectra due to the short length of realization. They are best expressed 
  in cycle 23, where large-scale field were more powerful.

  Periods less than 100 days are  clearly visible at 23 and 24 cycles in both types of groups. 
  They appear better in the period of the cycle maximum. Separately should be said about the 
  period of 113 days, which is shown only in the more powerful 23 cycle and only in flaring 
  (that means apparently larger) groups. One can make a conservative assumption that it is
  related to the characteristic time of the existence of large activity complexes 4-5 solar rotations.
  
 Therefore we can suggest that the classical dynamo generates a distributed large-scale field, while the additional dynamic process gives rise to the appearance of active regions. This second process does not end with the formation of the active region, but is expressed in a variety of relative movements of sunspots and the emergence of a magnetic flux. Thus, the flare is a natural extension of the process of formation of the active region. In the active regions where the second process has stopped, the active regions are formed less effectively and retain the statistical properties of the large-scale field.

  %
   \begin{acks}
     All flaring and non-flaring AR data used in this study were taken from Space 
     Weather Prediction Center (SWPC). The wavelet software was provided by C. 
     Torrence and G. Compo, and is available at http://paos.colorado.edu/research/wavelets/.
     The MTM analysis software was downloaded from http://research.atmos.ucla.edu/.
     This study was supported by the Scientific and Technical Council of Turkey by 
     the Project of 115F031. V. Yurchyshyn acknowledges support from AFOSR FA9550-15-1-0322 and NSF AST-1614457 grants and KASI. JPRO acknowledges the International Space Science Institute in Bern
     (Switzerland) for a ”visitor scientist” grant.  
   \end{acks}

   %
   %
   \bibliographystyle{spr-mp-sola}
   \bibliography{solphnew} 

   \IfFileExists{\jobname.bbl}{} {\typeout{}
    \typeout{****************************************************}
    \typeout{****************************************************}
    \typeout{** Please run "bibtex \jobname" to obtain} \typeout{**
      the bibliography and then re-run LaTeX} \typeout{** twice to fix
      the references !}
    \typeout{****************************************************}
    \typeout{****************************************************}
    \typeout{}} 

\end{article} 

\end{document}